\documentclass[12pt]{article}
\usepackage{latexsym,amsmath,amssymb,}
\usepackage{graphicx}

\def\be{\begin{equation}}
\def\ee{\end{equation}}
\def\beq{\begin{eqnarray}}
\def\eeq{\end{eqnarray}}

\def\D{\mathcal{D}}
\def\bes{\begin{eqnarray}}
\def\ees{\end{eqnarray}}

\newlength{\sizeonefig}
\newlength{\sizetwofig}
\newcommand{\nn}{\nonumber\\}

\begin{document}

\title{Quasiperiodic oscillations and  Tomimatsu-Sato $\delta=2$ space-time }

\author{Ivan Zh. Stefanov$^{1}$\thanks{E-mail: izhivkov@tu-sofia.bg}  \,\,\,, \,\, Galin G. Gyulchev$^{2}$\thanks{E-mail: gyulchev@phys.uni-sofia.bg}  \,\,\,, \,\,
Stoytcho S. Yazadjiev$^{3}$\thanks{E-mail: yazad@phys.uni-sofia.bg}\\\\
{\footnotesize{$^{1}$ Department of Applied Physics, Technical University of Sofia}}\\
{\footnotesize 8, St. Kliment Ohridski Blvd., 1000 Sofia, Bulgaria}\\\\[-3.mm]
{\footnotesize $^{2}$ Department of Physics, Biophysics and Roentgenology, }\\
{\footnotesize  Faculty of Medicine, St.Kliment Ohridski University of Sofia}\\
{\footnotesize  1, Kozyak Str., 1407 Sofia, Bulgaria}\\\\[-3.mm]
{\footnotesize $^{3}$ Department of Theoretical Physics, Faculty of Physics, St.Kliment Ohridski University of Sofia}\\
{\footnotesize  5, James Bourchier Blvd., 1164 Sofia, Bulgaria }}



\maketitle

\begin{abstract}
We model the spacetime of  low-mass X-ray binaries with the Tomimatsu-Sato $\delta=2$ (TS2) metric and study the properties of the orbital and the epicyclic frequencies. The numerical analysis shows that the properties of the characteristic frequencies of oscillation do not differ qualitatively from those of the Kerr black hole. Estimates for the angular momenta of the three stellar mass black hole candidates GRO 1655-40, XTE 1550-564 and GRS 1915+105 are made with the application of the nonlinear resonance model. We find agreement between the predictions  based on the $3:2$ nonlinear resonance model for a TS2 background and the current estimates found in the literature.

\end{abstract}

\noindent PACS numbers: 

\maketitle

\section{Introduction}
Quasiperiodic oscillations (QPO) in the X-ray flux from low-mass (galactic) binaries has attracted considerable research interest recently due to their potential to be applied as a tool for probing the space-time around black holes and massive stars and testing gravity in strong field regime \cite{KlisReview, PsaltisProbes, PsaltisNormanCited, DeDeoPsaltisSTT, JohansenPsaltis}.  One of the major possible application of QPOs which makes them so attractive is the measurement of the angular momentum of the central object in X-ray binaries. There are very few methods for this measurement \cite{Sreekumar}: continuum fitting of thermal spectra \cite{McClintock1, McClintock2, McClintock3}, determining the shape of the gravitationally redshifted wing of Fe line\cite{FeLine} and QPOs \cite{Mukhopadhyay}. Three types of QPO are usually observed: low-frequency QPOs (few tens of Hertz), intermediate frequency QPOs and high frequency QPOs (up to kilo Hertz). While the former two types of QPOs are believed to have astrophysical origin, i.e. they are related with the physics of the accretion disc, the high frequency QPOs are believed to depend on/reflect the structure of space-time in the vicinity of the central object --  a black holes or a neutron star. Some properties of the QPOs support that idea: high-frequency QPOs depend very weakly on the X-ray flux and in several sources the frequencies of the QPOs occur in pairs whose ratio is $3:2$. The physical mechanism of the production of these QPOs is not know. One of the major hypothesis is proposed by the nonlinear resonance model (NRM) according to which QPOs are related to the characteristic frequencies of oscillation -- epicyclic and orbital -- of test particles orbiting around the central object and the appearance of nonlinear resonances between them. This idea has initially been proposed by Aliev and Galtsov \cite{AlievGaltsov1, AlievGaltsov2}. It has reappeared and further elaborated in the works of Abramowicz, Kluzniak and collaborators \cite{AbramowiczInterpreting,AbramowiczSpinEstimate, AbramowiczTheory}. Simplicity makes the nonlinear resonance model  very attractive, however, it has some significant deficiencies \cite{RebuscoDifficulties}. The model does not propose an excitation mechanism for the QPOs. The origin of the coupling between the different characteristic frequencies is not clear.
The effect from astrophysical complication such as turbulence flow and magnetic instability have not been assessed and cannot be simply included in the model. All QPO models are purely dynamical and do not concern the emission mechanism of X-rays. A major difficulty of the NRM is the dissonance between its predictions for the angular momenta of the observed black holes candidates and the measurements based on spectral continuum fitting \cite{SpinProblem, RebuscoDifficulties}. Non of the different versions of the NRM can explain the observed angular momenta of the three black-hole candidates GRO 1655-40, XTE 1550-564 and GRS 1915+105.  In the current paper we will address only the last one of the problems. According to Bambi \cite{Bambi2012} the dissonance between the predictions of the NRM and the measurements for the angular momenta mean that one of the following four possibilities must be true: $i)$ The NRM is wrong $ii)$ continuum-fitting method does not provide reliable estimates of the angular momenta  $iii)$ both techniques do not work correctly  $iv)$ the central objects of the observed stellar-mass black hole candidates are not Kerr black holes. Here we will consider only the fourth possibility.

The effect of deviations from Kerr space-time on the QPOs in the frame of RNM has already been studied in several papers \cite{AlievKerr}, \cite{Bambi2012}, \cite{JohansenPsaltis}, \cite{braneworld}. In all of these cases the deviation from Kerr space-time is described by an additional continuous free parameter ($\epsilon$, tidal charge, etc.) which leads to degeneracy of the results in the sense that for fixed of the central object mass and value of the deviation parameter there is a whole interval of values of the angular momentum which are admissible by the observations. Due to this degeneracy, in order to be able to determine the value of the deviation parameter we need independent measurements of the mass and the angular momentum. One of the few exact solutions that describe the exterior field of a stationary rotating axially symmetric objects is the Tomimatsu-Sato solution. The deviation parameter $\delta$ in this solution takes discrete values ($\delta=1$ corresponding to Kerr solution $\delta=2$ corresponding to an exotic object) which allows the degeneracy to be avoided . The aim of the present paper is to apply the NRM to the space-time of Tomimatsu-Sato with $\delta=2$ (TS2), to study the properties of the characteristic frequencies and to check if the dissonance problem can be resolved in such background. The paper is organized as follows. The Tomimatsu-Sato spacetime is briefly presented in Section \ref{TS2}. The formulas that we use for  the calculation of the epicyclic frequencies are obtained in Appendix A and applied in Section \ref{freqs}.  In Section \ref{NRM} NRM model is sketched. The results for the estimates of the angular momenta of the three black hole candidates are in Section \ref{estimates}. The sum-up of the results is in Section \ref{SumUp}

\section{Tomimatsu-Sato space-time}\label{TS2}
The Kerr-Tomimatsu-Sato family is usually presented in the canonical Weyl-Papapetrou form of the metric for stationary, axisymmetric space-time\cite{Kra80}
\begin{equation}
 ds^2=-f(dt-\omega d\phi)^2+f^{-1}[e^{2\gamma}(d\rho^2+dz^2)+\rho^2d\phi^2],
\end{equation}
where \cite{TS73}
\begin{equation}
f=\frac{A}{B} ,\quad \omega=\frac{2mq}{A}(1-y^2)C ,
\quad e^{2\gamma}=\frac{A}{p^{2\delta}(x^2-y^2)^{\delta^2}}.
\end{equation}
The functions $A,B$ and $C$ are polynomials of the prolate
spheroidal coordinates $x$ and $y$ defined by
\begin{equation}
  \rho=\sigma\sqrt{(x^2-1)(1-y^2)},\quad z=\sigma xy.
\end{equation}
In explicit form for the Kerr space-time $ \delta=1 $ these polynomials are
\begin{eqnarray}
 A=p^2(x^2-1)-q^2(1-y^2),\quad B=(px+1)^2+q^2y^2,
\quad C=-px-1.
\end{eqnarray}
In the Tomimatsu-Sato case $\delta=2$ , $A,B$
and $C$ are given by
\begin{eqnarray}
&&A=p^4(x^2-1)^4+q^4(1-y^2)^4\nonumber\\
  &&\qquad  -2p^2q^2(x^2-1)(1-y^2)\{2(x^2-1)^2+2(1-y^2)^2+3(x^2-1)(1-y^2)\}\nonumber\\
 &&B=\{p^2(x^2+1)(x^2-1)-q^2(y^2+1)(1-y^2)+2px(x^2-1)\}^2\nn
  &&\qquad  +4q^2y^2\{px(x^2-1)+(px+1)(1-y^2)\}^2\\
 &&C=-p^3x(x^2-1)\{2(x^2+1)(x^2-1)+(x^2+3)(1-y^2)\}\nn
  &&\qquad  -p^2(x^2-1)\{4x^2(x^2-1)+(3x^2+1)(1-y^2)\}+q^2(px+1)(1-y^2)^3\nonumber.
\end{eqnarray}
The Kerr-TS family has three free parameters $\sigma,p$ and $q$ which are related to the mass $M$ and the angular momentum
$J$ of the central object as
\begin{equation}
 p^2+q^2=1,\quad \sigma=\frac{Mp}{\delta},\quad J=M^2q.\label{TS_params}
\end{equation}
The properties of the TS2 space-time have been studied in several papers \cite{GibbonsRussellClark, Kodama, TS_holography, Manko}. A simple sketch of the structure of TS2 can be found in \cite{BambiTS2}. We will just briefly mention its main properties. The TS2 space-time has two spherically symmetric Killing horizons symmetrically distributed with respect to the equatorial plane at $\rho = 0$ and $z = ±\sigma$ ($x = 1$ and $y = ±1$) connected by a rod-like conical singularity which carries all the gravitational mass. The Killing horizons are everywhere regular except at the points of connection with the rod-like singularity. In the equatorial plane there is a ring singularity at the roots of $B(x, y = 0) = 0$ with zero Komar mass and infinite circumference. The  metric component $g_{\phi\phi}$ diverge on the ring singularity and is negative between the ring singularity and the central rod-like singularity. The latter means that there is causality violation and closed time-like curves in that region.

The formula for the quadrupole momentum of the Kerr-Tomimatsu-Sato family has been correctly derived in \cite{Manko}
\begin{equation}
Q=-M^3\left(\frac{\delta^2-1}{3\delta^2}p^2+q^2\right).  \label{Q_TS}
\end{equation}

Due to its peculiar causal structure the TS2 space-time is discarded as a candidate to describe the final stage of gravitational collapse. The more appealing physical interpretation is the one given by Gibbons and Russell-Clark \cite{GibbonsRussellClark} that it describes the exterior field of a specific star whose prolateness is bigger than that of the Kerr black hole. The stability of TS2 has not been studied up to now.

\section{Characteristic epicyclic and orbital frequencies}\label{freqs}
The epicyclic frequencies of relativistic compact objects to our knowledge have been derived for the first time in \cite{AlievGaltsov1}. General formulas for the vertical and radial epicyclic frequencies can be found in \cite{OrbitsNovikovManko} and \cite{Pappas}. In  appendix \ref{appendix} of the current paper we present a slightly different derivation. In this derivation the decoupling of the vertical and the radial modes is not assumed but rather follows from the reflectional symmetry of the background metric (For more details on reflectional symmetry see \cite{MeinelNeugebauer}.).
The vertical and radial epicyclic frequencies, respectively, are  $\nu_{\rho}^2=\D^{\rho}_{ \,\,\,\, {\rho}}/(2\pi)$ and $\nu_{z}^2=\D^{z}_{ \,\,\,\, {z}}/(2\pi)$. The matrix $\D$ is defined in formulas (\ref{eq_final}) and (\ref{m_eq}) form Appendix \ref{appendix}. For the particular background metric, the TS2 spacetime, the off-diagonal components vanish, $\D^{z}_{ \,\,\,\, {\rho}}=0$ and $\D^{\rho}_{ \,\,\,\, {z}}=0$.
For the Keplerian frequency $\Omega_K$ we apply the formula
\be
\Omega_{\rm K} = \frac{d\phi}{dt} =
\frac{- \partial_{\rho} g_{t\phi}
\pm \sqrt{\left(\partial_{\rho} g_{t\phi}\right)^2
- \left(\partial_{\rho} g_{tt}\right) \left(\partial_{\rho}
g_{\phi\phi}\right)}}{\partial_{\rho} g_{\phi\phi}}.
\ee
Plus ``$+$'' refers to prograde orbits and minus ``$-$'', to retrograde.
For the calculation of the characteristic frequencies of the TS2 space-time we apply a package linear algebra in \textit{Maple}. The formulas are too cumbersome to be displayed explicitly.

In some of the particular cases in which the resonance model was applied that we cited in the Introduction it was found that the characteristic frequencies may have rather odd behavior qualitatively different from the case of Kerr black hole. Vertically unstable modes have been reported in \cite{braneworld, Bambi2012}, multiple (rather than one) extrema of the radial frequencies -- in \cite{Kerr_naked_sing, braneworld}, violation of the $\nu_k>\nu_{\theta}>\nu_r$ inequality, where $\nu_{\theta}$ and $\nu_r$ are the frequencies of the vertical and the radial modes respectively, has been shown to exist in \cite{Bambi2012, Kerr_naked_sing, braneworld}, absence of radially unstable modes has been found in \cite{braneworld}.

\begin{figure}[htbp]
\center
\includegraphics[width=0.47\textwidth]{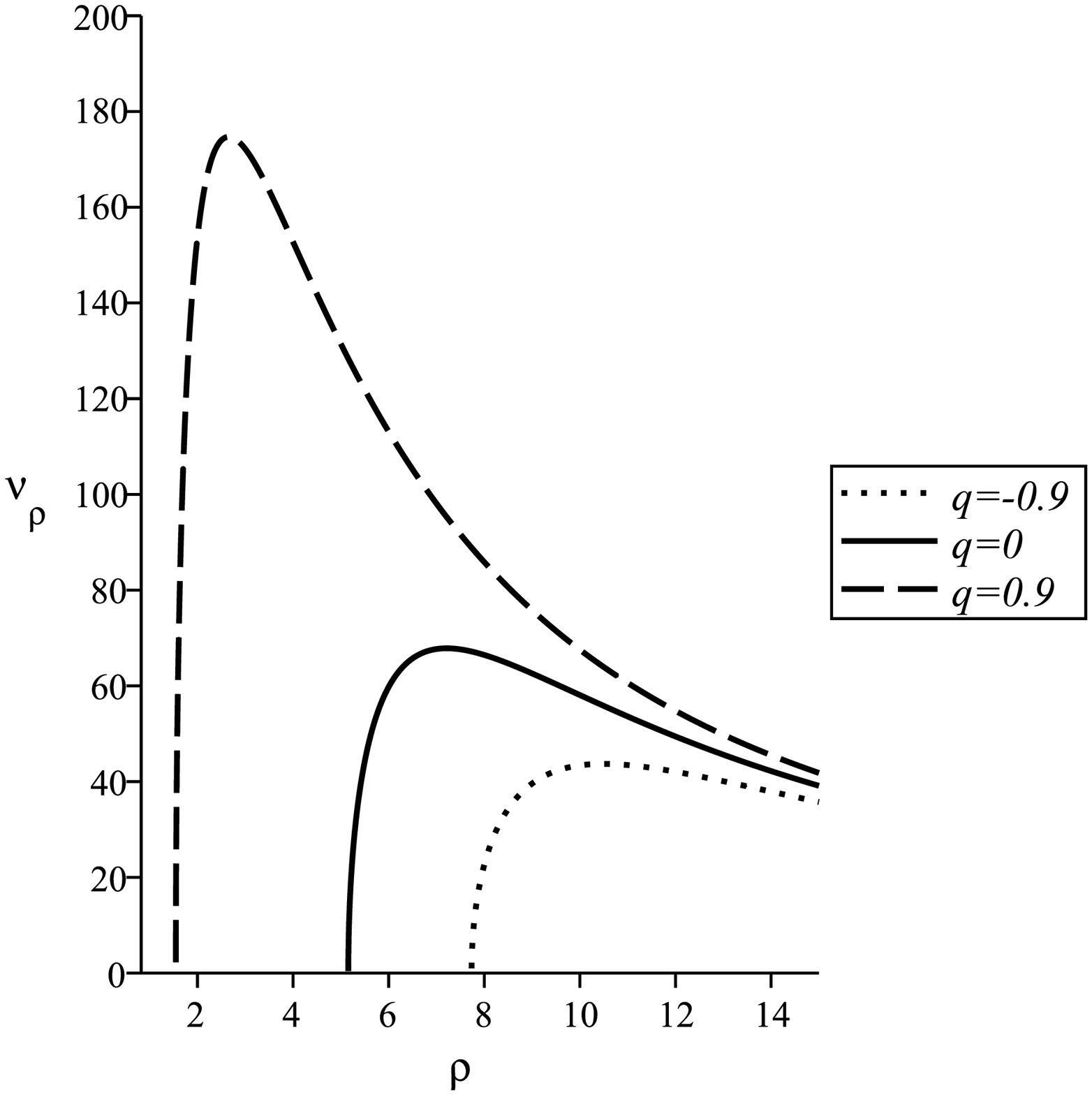}
\includegraphics[width=0.47\textwidth]{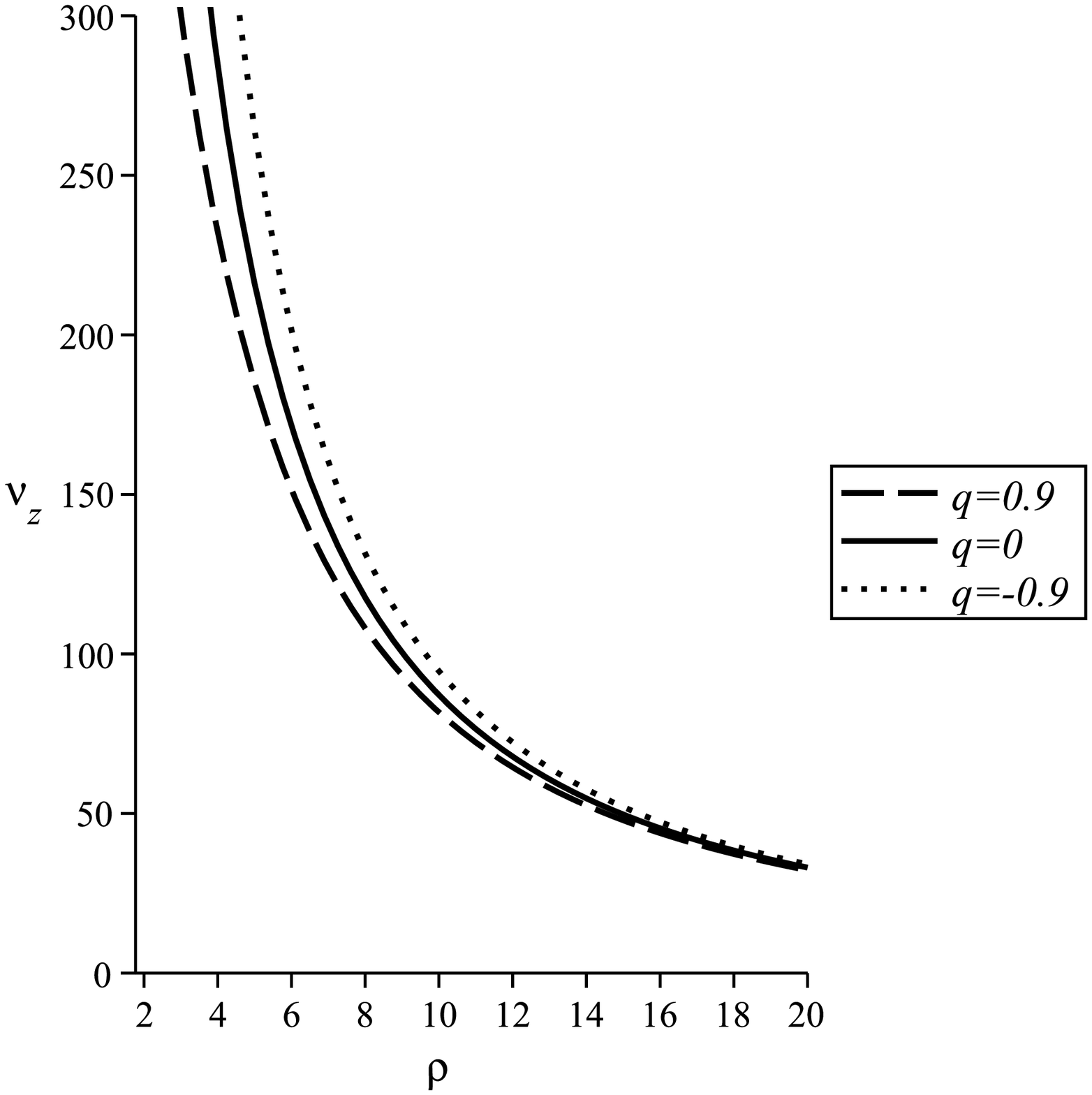}
\caption{The radial and the vertical epicyclic frequencies,  $\nu_{\rho}$ and $\nu_z$, for $q=-0.9; \,\,\, 0; \,\,\, 0.9$.} \label{epicyclic}
\end{figure}

\begin{figure}[htbp]
\center
\includegraphics[width=0.47\textwidth]{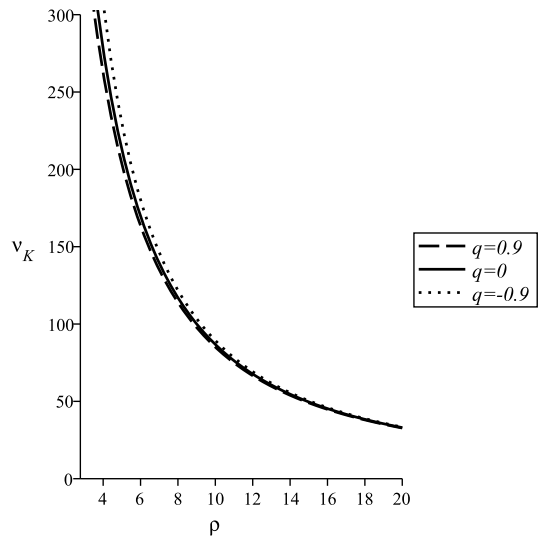}
\caption{The orbital frequency $\nu_K$ for $q=-0.9; 0; 0.9$.} \label{orbital}
\end{figure}

\begin{figure}[htbp]
\center
\includegraphics[width=0.5\textwidth]{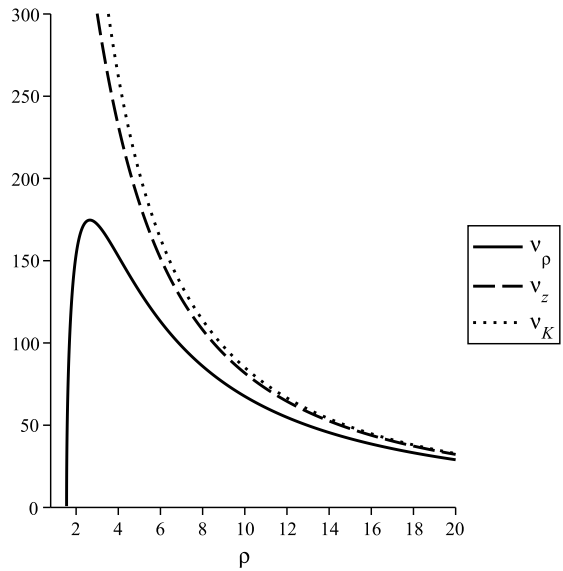}
\caption{The radial and the vertical epicyclic frequencies $\nu_{\rho}$ and $\nu_z$ and the orbital frequency $\nu_{K}$ for $q=0.9$.} \label{all_freq}
\end{figure}

\begin{figure}[htbp]
\center
\includegraphics[width=0.5\textwidth]{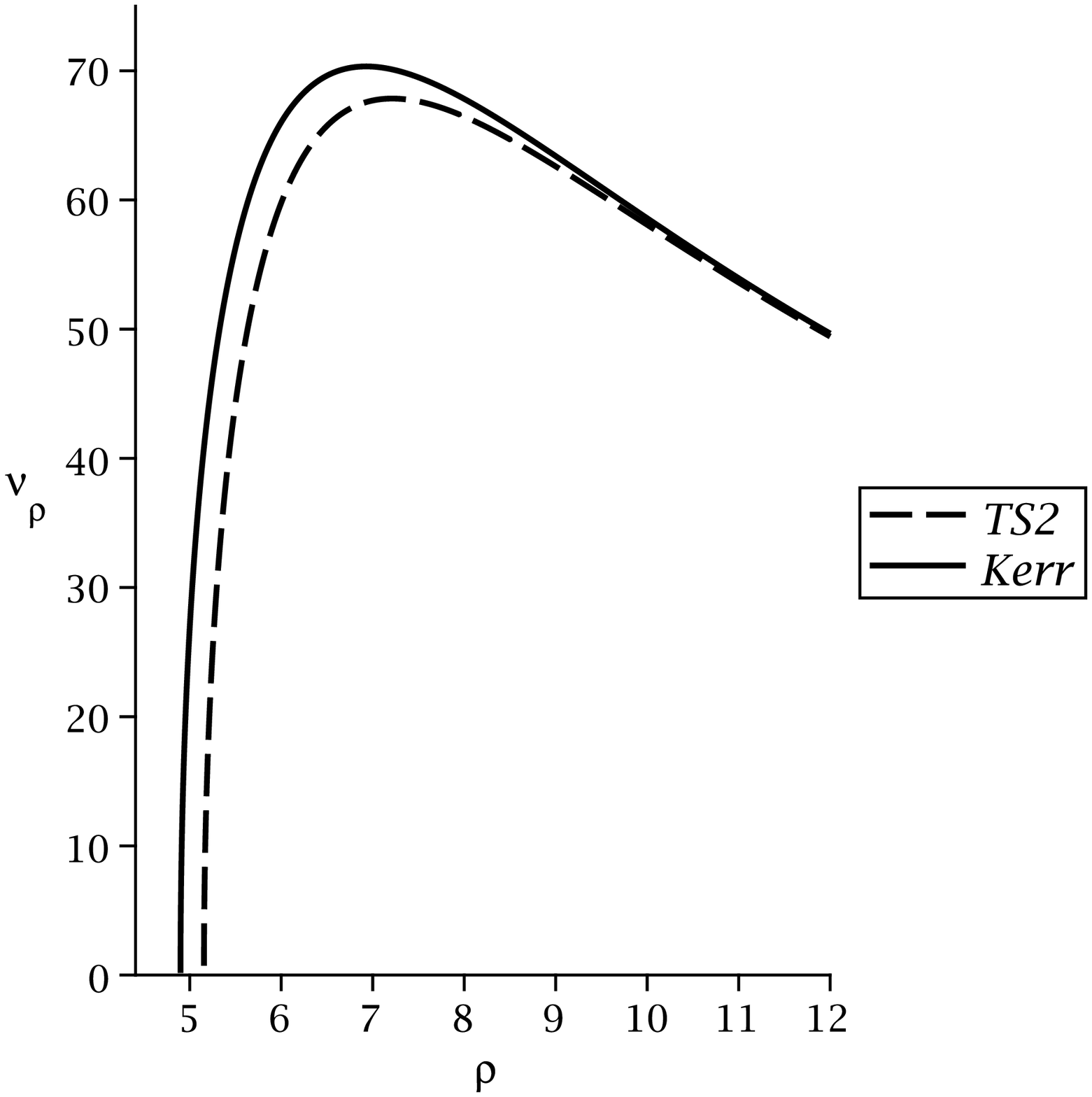}
\caption{A comparison between the Kerr and TS2 radial frequencies $\nu_{\rho}$ for $q=0$.} \label{Kerr_TS2}
\end{figure}

Unlike the cases discussed above there are no peculiarities in the properties of the epicyclic frequencies of the TS2 space-time. Qualitatively
their behavior is the same as in the case of Kerr black hole. The radial frequency $\nu_{\rm \rho}$ has only one extremum (maximum).
It increases as the singularity is approached, passes through a maximum and becomes zero
at $\rho=\rho_{ISCO}$\footnote{ISCO stands for innermost stable circular orbit.} before the singularity is reached. The frequency $\nu_{\rm \rho}$ is presented on the left panel of figure (\ref{epicyclic}) for $q=-0.9; 0; 0.9$. The same behavior is observed for the whole interval of admissible values of $q$. As we can see from the representative cases shown on the figure for positive angular momenta of the central object the frequencies are higher than for the negative angular momenta. The vertical epicyclic frequency $\nu_{z}$ and the orbital frequency $\nu_K$ increase monotonically as the singularity is approached (See the right panel of figure (\ref{epicyclic}) and figure (\ref{orbital}) ). For these two frequencies the effect of $q$ is opposite in comparison to the radial frequency case. Negative angular momenta invoke higher frequencies while positive angular momenta -- lower frequencies. All three frequencies $\nu_{\rho}$, $\nu_z$ and $\nu_K$ are displayed on figure (\ref{all_freq}) for $q=0.9$. For corotating orbits  $\nu_{\rho}<\nu_z<\nu_K$. As we have already mentioned the TS2 space-time is a significant deformation of Kerr space-time. The properties of these two space-times differ significantly even when $q=0$. On figure (\ref{Kerr_TS2}) $\nu_{\rho}$ for $q=0$ is shown and as it can be seen  $\nu_{\rho}$ is lower in the TS2 case.
The radius of the TS2 ISCO is higher than the one of Kerr. The differences of $\nu_z$ and $\nu_K$ in the two cases are indistinguishable on the graphics.

\section{Nonlinear resonance model for high frequency QPOs}\label{NRM}
According to the NRM resonances occur at the orbits for which the relation
\be
m\nu_{\rho}(q,r)=n\nu_z(q,r)\label{resonance_eq}
\ee
is satisfied for integer ``$m$'' and ``$n$''. The values of ``$m$'' and ``$n$'' vary in the different resonance models. When ``$q$'' is fixed (\ref{resonance_eq}) can be solved for the radii of the resonance orbits $r_{nm}$. For the observed twin peak high-frequency QPOs the upper $\nu_U$ and the lower $\nu_L$ frequency are in $3:2$ ratio so within the model they are expressed in the following way
\be
\nu_U=m_1\nu_{\rho}+m_2\nu_z, \quad\quad\nu_L=n_1\nu_{\rho}+n_2\nu_z, \label{upper_lower}
\ee
where the epicyclic frequencies are evaluated at the resonance orbits.
The small integers $\{m_1, m_2, n_1, n_2\}$ are chosen appropriately so the $\nu_U:\nu_L=3:2$ ratio is satisfied.
For the forced oscillation model $(m,n)=(3,2)$. In the parametric resonance model either $(m,n)=(3,1)$ or $(m,n)=(2,1)$.

In the keplerian resonance model the resonance condition is, respectively,
\be
m\nu_{\rho}(q,r)=n\nu_K(q,r),\label{resonance_eq_K}
\ee
where $(m,n)=(3,2);(3,1);(2,1)$. The upper and the lower twin peak QPO frequencies are
\be
\nu_U=m_1\nu_{\rho}+m_2\nu_K, \quad\quad\nu_L=n_1\nu_{\rho}+n_2\nu_K. \label{upper_lower_K}
\ee
Again the radial epicyclic frequency and the orbital frequency are evaluated at the resonance orbits.

\section{Estimates for the angular momenta of black-hole candidates in microquasars}\label{estimates}

In this section we apply different nonlinear resonance models to estimate the angular momenta of the central object in three microquasars GRO 1655-40, XTE 1550-564 and GRS 1915+105. For these black holes candidates we have also dynamical measurement of the mass. For all three of them two high-frequency QPOs  have been observed and the ratio between the upper $\nu_U$ and the lower frequency $\nu_L$ is $3:2$  (See Table. \ref{table_mass_freq} ). Two high-frequency QPOs have been observed also in the X-ray binary H 1743-322, however, there is no measurement for its mass.

\begin{table}
\begin{center}
\begin{tabular}{ l  l  l  l}
\\
\hline
Model  &GRO 1655--40 &XTE 1550--564  &GRS 1915+105\\
\hline
\hline
$M/M_\odot$      &  $6.30 \pm 0.27$   &  $9.1 \pm 0.6$  &  $14.0 \pm 4.4$\\
\hline
$\nu_{U}$, [Hz]  &  $450 \pm 3$       &   $276 \pm 3$   &   $168 \pm 3$\\
\hline
$\nu_{L}$, [Hz]  &  $300 \pm 5$       &   $184 \pm 5$   &   $113 \pm 5$\\
\hline
     \end{tabular}
     \end{center}
\caption{ Stellar-mass BH candidates in microquasars with a measurement of the
mass and two observed high-frequency QPOs \cite{stro01, rem02, mcc}.}
\label{table_mass_freq}
\end{table}
The results from the numerical calculation for the estimates of the angular momenta of the microquasars based on the models in which nonlinear resonance occurs between the radial and vertical epicyclic frequencies are presented in Table. \ref{table} and for resonance between the radial epicyclic frequency and the orbital frequancy -- in Table. \ref{table_Kepler}.
\begin{table}
\begin{center}
\begin{tabular}{ l  l  l  l }
\\
\hline
Model  &GRO 1655--40 &XTE 1550--564  &GRS 1915+105\\
\hline
\hline
~~3:2 &~$0.63~\div~0.68$ &~$0.41~\div~0.55$ & $-0.11~\div~0.68$\\
\hline
~~2:1 &~$0.40~\div~0.49$ &~$0.21~\div~0.38$ &  $-0.40~\div~0.55$\\
\hline
~~3:1 &~$0.58~\div~0.66$ &~$0.42~\div~0.57$ &  $-0.12~\div~0.71$\\
\noalign{\smallskip}
\hline
     \end{tabular}
     \end{center}
\caption{ Angular momentum estimates for the three microquasars  GRO 1655-40, XTE 1550-564 and GRS 1915+105 from the models in which the resonance occurs between the radial and vertical epicyclic frequencies.}
\label{table}
\end{table}
\begin{table}
\begin{center}
\begin{tabular}{ l  l  l  l }
\\
\hline
Model  &GRO 1655--40 &XTE 1550--564  &GRS 1915+105\\
\hline
\hline
~~3:2 &~$0.62~\div~0.71$ &~$0.45~\div~0.61$ & $-0.14~\div~0.76$\\
\hline
~~2:1 &~$0.40~\div~0.49$ &~$0.21~\div~0.38$ &  $-0.40~\div~0.55$\\
\hline
~~3:1 &~$0.54~\div~0.61$ &~$0.39~\div~0.52$ &  $-0.10~\div~0.65$\\
\noalign{\smallskip}
\hline
     \end{tabular}
     \end{center}
\caption{Angular momentum estimates for the three microquasars  GRO 1655-40, XTE 1550-564 and GRS 1915+105 from the models in which the resonance occurs between the radial epicyclic frequency and the orbital (Keplerian) frequency.}
\label{table_Kepler}
\end{table}
In general we can say that when the central object of the microquasar is modeled by the central object of TS2 space-time the estimates for the angular momenta are lower than in the case when the central object is modeled by a Kerr black hole. The estimates in the latter case can be found in  \cite{AbramowiczSpinEstimate}. One of the most significant problems of the nonlinear resonance model is the discrepancy between the estimates for the angular momenta based on the observed QPOs and the estimates obtained thought analysis of the observed thermal spectra of the objects' accretion disks. In non of the resonance models there is agreement between QPOs and thermal spectra for all three microquasars. For example, the measured value of angular momentum of GRS 1915+105 coming from the continuum-fitting method is in agreement with the $3:2$ epicyclic resonance model but the value predicted by the same resonance model for GRO 1655-40 does not match the spectral fitting estimate.  It would be interesting to know whether such discrepancy is still present when TS2 solution is used to describe the space-time in the vicinity of the microquasar.
In order to be able to legitimately compare between the estimates of the resonance model of QPOs and those of the thermal spectra fitting the experimental data in the later case should be reanalyzed with TS2 background. In the absence of such analysis we could compare the estimates of the resonance model for TS2 with the estimates of the spectral continuum-fitting and other methods for Kerr. The continuumm-fitting method \cite{McClintock2} gives the following values for GRO 1655-40, $a\in(0.65\div0.75)$. The angular momentum of XTE 1550-564 has been estimated in \cite{McClintock3} to be $a\in(0.29\div0.52)$. These values have been obtained through the combination of the results from the continuum-fitting method and the Fe-line method.  In literature two conflicting estimates for GRS 1915+105 can be found (See also \cite{Strohmayer}). The first one is given in \cite{McClintock1} -- $a>0.98$. The estimate given in \cite{GRS_alternative} is $a\sim0.7$. From tables \ref{table} and \ref{table_Kepler} we can see that the $3:2$ Keplerian resonance is consistent with these values for all three microquasars if the alternative estimate for the angular momentum of GRS 1915+105 \cite{GRS_alternative} is chosen.

\section{Sum-up}\label{SumUp}
In the current paper we have studied the properties of the three characteristic frequencies of a particle propagating alone a circular orbit in the in the equatorial plane of Tomimatsu-Sato space-time -- the radial $\nu_{\rho}$ and the vertical $\nu_z$ epicyclic frequencies and the orbital  (Keplerian) frequency $\nu_{K}$. For the calculation of the frequencies we have followed a variation of the method proposed \cite{AlievGaltsov1}. The formulas are presented in appendix A of the current paper. The numerical analysis shows that for the whole interval of admissible values of the free parameter $q$ corresponding to angular momentum per unit mass the quantitative behavior of the frequencies are the same as in the case of Kerr black hole.
We have also applied the resonance mode to estimate the angular momenta of three microquasars GRO 1655-40, XTE 1550-564 and GRS 1915+105. The central object here is modeled by the central object in the Tomimatsu-Sato $\delta=2$ space-time. The results are presented in Table. \ref{table} and Table. \ref{table_Kepler}. The estimated  angular mementa of the three microquasars are considerably lower than in the case when the central object is modeled by a Kerr black hole.

The $3:2$ keplerian NRM seems to be  in agreement with the some of the current estimates for the angular momenta of three black hole candidates. However, future studies should take into account that the  continuum-fitting method is model dependent and in the cases cited here it was supposed that the central object is a Kerr black hole.

In conclusion, the results presented here do not exclude the Tomimatsu-Sato $\delta=2$ space-time as a model of the space-time of microquasars.
\appendix
\section{Derivation of the formula for the epicyclic frequencies}\label{appendix}

\be
\ddot{x}^\mu+\Gamma^\mu_{\,\,\,\alpha\beta}\dot{x}^\alpha\dot{x}^\beta=0\label{geod_eq}
\ee
\be
\delta\ddot{x}^\mu+2\Gamma^\mu_{\,\,\,\beta\sigma}\dot{x}^\beta\delta \dot{ x}^\sigma+\Gamma^\mu_{\,\,\,\alpha\beta,\sigma}\dot{x}^\alpha\dot{x}^\beta\delta  x^\sigma=0\label{geod_pert_eq}
\ee
We are interested in circular orbits in space-times with two Killing vectors $\xi_t$ and $\xi_{\phi}$ --  stationary (or static in particular) and axially symmetric . Due to the invariance of the space-time with respect to translations in time and rotations with respect to an axis the law of conservation of energy and the law of conservation of angular momentum hold for point particles. We will use adapted coordinates in which the two Killing vectors have the form $\xi_{t}=\partial/\partial t$ and $\xi_{\phi}=\partial/\partial \phi$. The solutions whose QPOs we will study are presented Weyl-Papapetrou ($t, \rho, z, \phi $) or Boyer-Lindquist ($ t, r, \theta, \phi $) coordinates. The derivation of the epicyclic frequencies presented here is valid also if prolate spheroidal coordinates are used to present an axially symmetric solution. By flat orbits here we mean orbits that remain in a plane and in particular in the equatorial plane. In the coordinate systems we have chosen these orbits are described as $\rho={\rm const}$ and $z={\rm const}$ or alternatively $r={\rm const}$ and $\theta={\rm const}$.  For convenience we will adopt the following notation. Small Latin letters $a, b, c, d, s, m, n....$ will be used to denote the $r$ and $\theta$ (respectively the $\rho$ and $z$) coordinates. Capital Latin letters $A, B, C, D, S, M, N....$ represent the other two coordinates, those related with the energy and angular momentum conservation laws,  $t$ and $\phi$ respectively. We will also term them cyclic coordinates since the lagrangian of point particles does not depend on them. Greek letter will denote all coordinates $\alpha, \beta, \gamma, \delta, \sigma, \mu, \nu...= t, r, \theta, \phi\,\,\,(t, \rho, z, \phi)  $. In that notation flat, circular orbits are expressed as $\dot{x}^{a}=0$ where the dot denotes the derivative with respect to the affine parameter of the orbits -- $\lambda$. Due to stationarity and axial symmetry for the metric we have $g_{\mu\nu}=g_{\mu\nu}(x^a)$. It can be proved that all quantities that are function of the $x^a$ coordinates only remain constant on the chosen type of orbits. Let $f=f(x^{a})$. Then
\be
\dot{f}={d f\over d \lambda} = f_{,a}\dot{x}^{a}=0,
\ee
since $\dot{x}^{a}=0$. In particular
\be
{d \over d \lambda}\left(\Gamma^\mu_{\,\,\,\alpha\beta}\right) =0.\label{Gamma_const}
\ee
For the particular case of flat, circular orbits $\ddot{x}^{a}=0$ while in general for the cyclic coordinates $\ddot{x}^{A}=0$ holds.
So
\be\ddot{x}^{\alpha}=0.\label{x_const}
\ee
Due to (\ref{Gamma_const}) and (\ref{x_const}),   (\ref{geod_pert_eq}) can be rewritten as
\be
{d \over d \lambda}\left(\delta\dot{x}^\mu+2\Gamma^\mu_{\,\,\,\beta\sigma}\dot{x}^\beta\delta  x^\sigma\right)=-\Gamma^\mu_{\,\,\,\alpha\beta,\sigma}\dot{x}^\alpha\dot{x}^\beta\delta  x^\sigma. \label{geod_pert_eq_const}
\ee
Taking into account that
\be
\Gamma^\mu_{\,\,\,\alpha\beta}={1\over2}g^{\mu\sigma}\left(\partial_{\alpha}g_{\beta\sigma}+\partial_{\beta}g_{\alpha\sigma}-
\partial_{\sigma}g_{\alpha\beta}\right)
\ee
we can easily show that some of the Christoffel symbols vanish.

\bes
\Gamma^{M}_{\,\,\,AB} =-{1\over2} g^{Ms}\partial_s g_{AB}=0, \quad\quad \Gamma^{m}_{\,\,\,AB} =-{1\over2} g^{ms}\partial_s g_{AB}\neq0; \notag\\
\Gamma^{M}_{\,\,\,Ab} ={1\over2} g^{MP}\partial_b g_{AP}\neq0, \quad\quad \Gamma^{m}_{\,\,\,Ab} ={1\over2} g^{mP}\partial_b g_{AP}=0.\label{Christoffel}
\ees
We have taken into account that $g_{AB}\neq0$ for any $A$ and $B$,  $g_{ab}\neq0$ for $a=b$,  $g_{Ab}=0=g_{aB}$ and for the inverse metric $g^{AB}\neq0$ for $A$ any $B$,  $g^{ab}\neq0$ for $a=b$,  $g^{Ab}=0=g_{aB}$.

The $M$ components of (\ref{geod_pert_eq}) simplifies considerably. From (\ref{geod_pert_eq_const}) and (\ref{Christoffel})
\be
{d \over d \lambda}\left(\delta\dot{x}^M+2\Gamma^M_{\,\,\,Bs}\dot{x}^B\delta x^s\right)=0
\ee
so
\be
\delta\dot{x}^M+2\Gamma^M_{\,\,\,Bs}\dot{x}^B\delta x^s={\rm const}=0.\label{M_comp}
\ee
Setting the constant to zero is equivalent to choosing the difference between initial phases of $\delta\dot{x}^M$ and $\delta x^\sigma$ to be $\pi/2$. Since $(d/dt) = \dot{t}^{-1}({d/d \lambda})$ from (\ref{M_comp}) it follows that
\be
{d \delta x^M\over dt}+2\left(\Gamma^M_{\,\,\,ts}+\Gamma^M_{\,\,\,\phi s}\Omega_{\rm K}\right)\delta x^s=0.\label{M_eq}
\ee
$\Omega_{\rm K}$ is the Keplerian frequency
\be
\Omega_{\rm K}={\dot{\phi}\over \dot{t}}={d \phi\over dt}={L\over E}.
\ee
$L$ and $E$ are the angular momentum and the energy of the particle, respectively.
Now let us obtain the equation that governs the $m$ component of the perturbations
\be
\delta\ddot{x}^m+2\Gamma^m_{\,\,\,BP}\dot{x}^B\delta \dot{ x}^P+\Gamma^m_{\,\,\,AB,s}\dot{x}^A\dot{x}^B\delta  x^s=0.
\ee
It contains both the $\delta x^m$ and the $\delta x^P$ perturbations. The $\delta x^P$ can be decoupled if (\ref{M_comp}) is used
\be
\delta\ddot{x}^m+2\Gamma^m_{\,\,\,BP}\dot{x}^B \left( -2\Gamma^P_{\,\,\,As}\dot{x}^A\delta x^s\right)+\Gamma^m_{\,\,\,AB,s}\dot{x}^A\dot{x}^B\delta  x^\sigma=0.
\ee
Finally,
\be
\delta\ddot{x}^m+F^m_{\,\,\,ABs}\dot{x}^A\dot{x}^B\delta  x^s=0,
\ee
where we have introduced the following matrix
\be
F^m_{\,\,\,ABs}=\Gamma^m_{\,\,\,AB,s} -4\Gamma^m_{\,\,\,AP}\Gamma^P_{\,\,\,Bs}.\label{F_matrix}
\ee
Dividing the upper equation in $\dot{t}$ we obtain
\be
{d^2 \delta x^m\over dt^2}+\left(F^m_{\,\,\,tts}+2F^m_{\,\,\,t\phi s}\Omega_{\rm K}+F^m_{\,\,\,\phi \phi s}\Omega_{\rm K}^2\right)\delta  x^s=0,\label{eq_final}
\ee
or
\be
{d^2 \delta x^m\over dt^2}+\D^m_{ \,\,\,\, s}\delta  x^s=0.\label{m_eq}
\ee
In the general case the horizontal ( $r$ or $\rho$) and the vertical ($\theta$ or $z$) perturbations are coupled. When the "elasticity"
matrix $\D$ is diagonal the epicyclic frequencies are
\be
\Omega_m^2=\D^m_{ \,\,\,\, m}, \quad\quad m= r, \theta\quad (\rho, z)\label{epicyclic_freq}.
\ee
No summation over the repeated index $m$ is implied.

\section*{Acknowledgements}
This work was partially supported by the Bulgarian National Science Fund under Grant No DMU 03/6.

%



\begin{thebibliography}{99}
\bibitem{KlisReview} M. van der Klis,  Astronomical Time Series, Eds. D. Maoz, A. Sternberg, and E.M. Leibowitz, 1997 (Dordrecht: Kluwer), p. 121, arXiv:astro-ph/9710016.
\bibitem{PsaltisProbes} D. Psaltis, Living Rev. Relativity 11 (2008), 9 , arXiv:0806.1531 [astro-ph].
\bibitem{PsaltisNormanCited} D. Psaltis, C. Norman, arXiv:astro-ph/0001391.
\bibitem{DeDeoPsaltisSTT} S. DeDeo, D. Psaltis,  American Astronomical Society, HEAD meeting $\sharp 8$, $\sharp19.04$; Bulletin of the American Astronomical Society, Vol. 36, p.944, arXiv:astro-ph/0405067.
\bibitem{JohansenPsaltis}T. Johannsen, D. Psaltis, Astrophys. J. \textbf{726}, 11 (2011), arXiv:1010.1000 [astro-ph.HE].
\bibitem{Sreekumar}B. Mukhopadhyay, D. Bhattacharya, P. Sreekumar,  IJMPD \textbf{21} No. 11, 1250086 (2012), arXiv:1210.2441 [astro-ph.HE].
\bibitem{McClintock1}[5] J. E. McClintock, R. Shafee, R. Narayan, R. A. Remil-
lard, S. W. Davis and L. -X. Li, Astrophys. J. \textbf{652}, 518 (2006), arXiv:astro-ph/0606076.
\bibitem{McClintock2}R. Shafee, J. E. McClintock, R. Narayan, S. W. Davis,
L. -X. Li and R. A. Remillard, Astrophys. J. \textbf{636}, L113 (2006).
\bibitem{McClintock3} J. F. Steiner, R. C. Reis, J. E. McClintock, R. Narayan,
R. A. Remillard, J. A. Orosz, L. Gou and A. C. Fabian
et al., Mon. Not. Roy. Astron. Soc. \textbf{416}, 941 (2011).
\bibitem{FeLine} J. L. Blum et al., Astrophys. J. \textbf{706},  60  ( 2009), arXiv:0909.5383 [astro-ph.HE].
\bibitem{Mukhopadhyay}  B. Mukhopadhyay, Astrophys. J. \textbf{694},387 (2009), arXiv:0811.2033 [astro-ph].

\bibitem{Bambi2012} C. Bambi, JCAP \textbf{1209}, 014 (2012), arXiv:1205.6348 [gr-qc].
\bibitem{AlievKerr} A. N. Aliev, G. D. Esmer, P. Talazan, arXiv:1205.2838 [gr-qc].
\bibitem{AlievGaltsov1} A. N. Aliev and D. V. Gal'tsov, Gen. Relat. Gravit. \textbf{13}, 899 (1981).
\bibitem{AlievGaltsov2} A. N. Aliev, D. V. Gal'tsov and V. I. Petukhov, Astr. Space Sci. \textbf{124}, 137 (1986).
\bibitem{AbramowiczInterpreting} M. A. Abramowicz, W. Kluzniak, AIP Conference Proceedings \textbf{714}, 21(2004), arXiv:astro-ph/0312396.
\bibitem{AbramowiczSpinEstimate} M. A. Abramowicz, W. Kluzniak, Z. Stuchlik, G. T\"{o}r\"{o}k,  arXiv:astro-ph/0401464.
\bibitem{AbramowiczTheory} G. Torok, M. A. Abramowicz, W. Kluzniak, Z. Stuchlik, AIP Conference Proceedings\textbf{861}, 786 (2006),arXiv:astro-ph/0603847.

\bibitem{SpinProblem} G. Torok, M. A. Abramowicz, Z. Stuchlik, E. Sramkova, Proceedings of XXVIth IAU General Assembly 2006
Proceedings IAU Symposium No. \textbf{240} (2006) B. Hartkopf, E. Guinan \& P. Harmanec, eds, arXiv:astro-ph/0610497 .
\bibitem{RebuscoDifficulties} P. Rebusco, New Astron. Rev. \textbf{51}, 855 {2008}, arXiv:0801.3658 [astro-ph].
\bibitem{braneworld}Z. Stuchlik, A. Kotrlova, \textit{Gen. Rel. Grav.} \textbf{41},1305 (2009),arXiv:0812.5066 [astro-ph].

\bibitem{Kra80} D.~Kramer, H.~Stephani, M.~MacCallum and E.~Herlt eds.:
	{\it Exact Solutions of Einstein's Field Equations}(Cambridge
	Univ. Press, Cambridge, 1980).
 \bibitem{TS73} Akira Tomimatsu and Humitaka Sato
Prog.\ Theor.\ Phys.\  {\bf 50}, 95 (1973).
\bibitem{GibbonsRussellClark}G. W. Gibbons and R. A. Russell-Clark, Phys. Rev. Lett. \textbf{30}, 398 (1973).
\bibitem{OrbitsNovikovManko}J. R. Gair, Ch. Li, I. Mandel, Phys. Rev. \textbf{D77}, 024035 (2008), arXiv:0708.0628 [gr-qc].
\bibitem{Pappas} G.Pappas, Mon. Not. R. Astron. Soc. \textbf{422}, 2581 (2012), arXiv:1201.6071 [astro-ph.HE].
\bibitem{MeinelNeugebauer} R. Meinel, G. Neugebauer, Class. Quant. Grav. \textbf{12}, 2045 (1995) arXiv:gr-qc/0302114.
\bibitem{Kodama}W. Hikida and H. Kodama, \emph{An Investigation of the Tomimatsu-Sato Spacetime},arXiv: gr-qc/0303094.
\bibitem{TS_holography} J. Gegenberg, H. Liu, S. S. Seahra and B. K. Tippett, Class. Quantum Grav. \textbf{28}, 085004 (2011), arXiv:1010.2803 [hep-th].
\bibitem{Manko}V.S. Manko,  Prog. Theor. Phys. \textbf{127}, 1057 (2012), arXiv:1110.6564 [gr-qc].
\bibitem{BambiTS2} C. Bambi, N. Yoshida, Class. Quant. Grav. \textbf{27}, 205006 (2010), arXiv:1004.3149 [gr-qc].
\bibitem{Kerr_naked_sing} G. Torok, Z. Stuchlik, \textit{A\&A} \textbf{437}, Issue 3,  775 (2005), arXiv:astro-ph/0502127.
\bibitem{stro01}
  T.~E.~Strohmayer,
  Astrophys.\ J.\  {\bf 552}, L49 (2001)
  [astro-ph/0104487].
\bibitem{rem02}
  R.~A.~Remillard, M.~P.~Muno, J.~E.~McClintock and J.~A.~Orosz,
  Astrophys.\ J.\  {\bf 580}, 1030 (2002)
  [astro-ph/0202305].
\bibitem{mcc}
  R.~A.~Remillard and J.~E.~McClintock,
  Ann.\ Rev.\ Astron.\ Astrophys.\  {\bf 44}, 49 (2006)
  [astro-ph/0606352].
\bibitem{Strohmayer}
  T.~E.~Strohmayer,
  Astrophys.\ J.\  {\bf 552}, L49 (2001)
  [astro-ph/0104487].
\bibitem{GRS_alternative} M. Middleton, Ch. Done, M. Gierlinski, Sh. Davis, MNRAS\textbf{ 373} Issue 3, 1004 (2006), arXiv:astro-ph/0601540.









\end{thebibliography}
\end{document}